# Sporadic Creutzfeldt－Jakob disease presenting with cerebral atrophy following traumatic brain injury mimicking hydrocephalus: a case report and literature review

Chun Zeng[a,1], Dezhu Gao[a,1], Liang Wu[a, *]

[a]Department of Neurosurgery, Beijing Tiantan Hospital, Capital Medical University, Beijing 100070, China

[1]These authors share the first authorship.

**Abstract**:

*Introduction*: Sporadic Creutzfeldt–Jakob disease (sCJD) is a rapidly progressive neurodegenerative disease without effective treatment that usually results in death within one year. The recently applied methods have improved the accuracy of the disease diagnosis and the specific radiological findings provide the necessary information for differential diagnosis.

*Research question*: The research is aimed to provide a different perspective on the development of CJD and associated literature review.

*Materials and methods*: The study presents a case who presented cognitive deficits, gait instability, and urinary and fecal incontinence suffered from traumatic brain injury eight months ago before admission with cerebral ventricle dilation on CT images. Furthermore, studies describe relevant cases are also included.

*Results*: The patient's symptoms got deteriorated. Further examinations, including 14-3-3 and tau proteins in the cerebrospinal fluid (CSF), MRI, and EEG, confirmed the patient's diagnosis of sCJD. He returned to the local hospital for the conservative treatment without effective medical intervention.

*Conclusion*: This case illustrates the diagnostic process of CJD and underscores the importance of distinguishing rare disorders from common conditions to achieve a comprehensive understanding of the disease.

Keywords: sporadic Creutzfeldt–Jakob disease, hydrocephalus, dementia

## 1. Introduction

Prion protein diseases are neurodegenerative diseases with the accumulation of misfolded prion protein ($PrP^{Sc}$) in humans and animals which result in a poor prognosis (du Plessis et al., 2008). Among prion protein diseases, Creutzfeldt-Jakob disease (CJD) is the most common type consisting of four major groups: sporadic (sCJD), genetic (gCJD), variant (vCJD), and iatrogenic (iCJD) (Uttley et al., 2020). sCJD is the predominant subtype of CJD, characterized by a median survival time of approximately one year after symptom onset (Orrú et al., 2014), with an estimated worldwide incidence of 1 case per million population annually (Wilson et al., 2019). Common symptoms of sCJD include rapidly progressive dementia, cerebellar ataxia, and myoclonus (Orrú et al., 2014). The hydrocephalus's triad, such as abnormal gait, memory retention, and fecal and urinary incontinence is usually identical to the appearance of sCJD. Special examinations and imaging results are useful for the differential diagnosis of sCJD.

In the following, we present a case whose initial CT imaging showed apparent local ventricular enlargement, along with clinical symptoms such as abnormal gait, dementia, and urinary incontinence. We first diagnosed the patient as hydrocephalus for the CT appearance and clinical symptoms. We decided to conduct a ventriculoperitoneal shunt (VP shunt) to relieve the symptoms at first. The abnormal presentations on MRI reminded us that further examinations were needed to confirm an exact diagnosis. Accurate diagnosis is crucial because it guides appropriate treatment and management strategies, which can significantly impact the patient's prognosis and quality of life. Further tests supported the diagnosis of sCJD rather than hydrocephalus. There is no effective method for the treatment of sCJD so further studies are needed to make great progress in this rare disorder treatment.

2. Case presentation

A 63-year-old man was admitted to our hospital with a history of traumatic brain injury 8 months ago, presenting with walking difficulty and cognitive deficit for 1 month. The patient was conscious and the movement was normal after the injury at first. After conservative therapy in the local hospital, he returned to normal living and continued to work. He began to appear unsteady gait with an increasingly worsening cognitive deficit 1 month before admission. Two weeks ago, he was unable to walk and exhibited dysarthria. The conditions deteriorated 1 week ago with the appearance of dysphagia and urinary incontinence. The terrible manifestations such as low spirits, difficulty falling asleep, decreased food intake, and weight loss also discomforted the patient. He was bedridden to the hospital with the stable vital signs. The patient was in a state of somnolence and reacted to the sharp pain. Due to the non-cooperation during physical examination, it was difficult to accurately examine the function of cranial nerves. The heart rate lasted at 137 beats per minute and SpO2 was 93% on the second day. The admission of esmolol to the patient relieved syndromes. He got a fever and transient loss of consciousness on the third day of hospitalization. The elevation of CRP and WBC indicated infectious signs. Positive anti-infection treatment through cefuroxime eased symptoms. Due to the exacerbation of dysphagia on that day, a nasogastric tube was inserted to facilitate the provision of nutrients. The ventricular enlargement on the CT imaging (Fig. 1B) combined with symptoms such as dementia and gait abnormality with the history of traumatic brain injury event strongly linked to the diagnosis of post-traumatic hydrocephalus. After the cognitive assessment on the 4th day, the CSF tap test was implemented to evaluate the accuracy of diagnosis and the probable efficacy after the V-P shunt. The routine lab test results of CSF as WBC, protein, lactate, and glucose were all normal. The CSF tap test result did not release the symptoms and the patient's situation worsened. The abnormal signal of right frontal, temporal, parietal, and insular lobes on MRI (Fig. 2) and a wide range of ring-like lesion hyperintensity in the right frontal cortex on DWI 17 days before admission had a strong link to the other neurological disease. After synthesizing the cognitive assessment results and imaging findings, the patient was deemed unlikely to be diagnosed with hydrocephalus. Compared with the former CT (Fig. 1A), the latter scan (Fig. 1B) showed focal dilatation of the right frontal horn, most likely due to localized scar-induced retraction causing neuronal death associated with brain atrophy rather than post-traumatic hydrocephalus. Therefore, further examinations were conducted to confirm the etiology and make the next treatment plan.

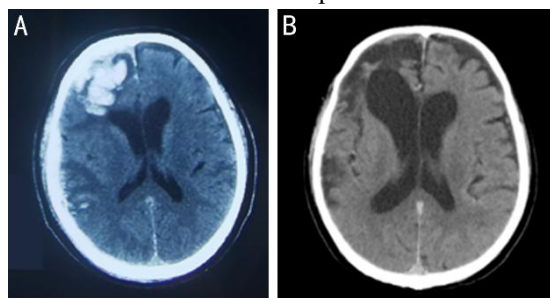

**Fig. 1** A CT scan comparison between the initial injury (A) and 8-month follow-up (B) condition.

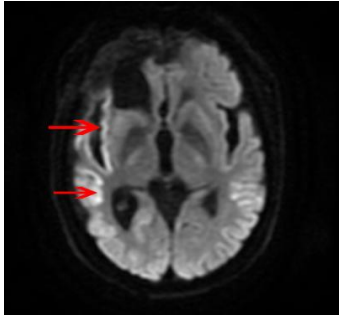

**Fig. 2** There was a wide range of ring-like hyperintense lesions in the right frontal cortex on DWI.

The CSF culture result showed that there was no infectious microbe in the patient's central nervous system. The negative result of NGS indicated that there was no infection in the patient's CSF (Dutra, et al., 2018). The antibody test in CSF was negative which poorly advocated the diagnosis of autoimmune encephalitis (AE) (Nagata, et al., 2023). Tissue-based assay (TBA) result showed no apparent positive signal which also decreased the possibility of AE (Weller, et al., 2018). The decreased proteins as $A\beta_{1-42}$, $A\beta_{1-40}$, and α-synuclein with the elevation of tau were the evidence of Alzheimer's disease (AD) (Shir, et al., 2022). The tau protein level increased more than fourfold, more supporting the diagnosis of CJD rather than AD. With the confirm of positive 14-3-3 protein result from CDC, the patient's accurate diagnosis was likely linked to the diagnosis of CJD (Hermann, et al., 2021). Further examination results also confirmed the diagnosis. EEG was abnormal for which there were significant diffuse δ waves in the bilateral frontal and temporal lobes with periodic sharp waves, and triphasic waves in the bilateral occipital and right middle posterior temporal lobes. Met homozygosity at PRNP codon 129 is considered to be the most susceptible genotype for CJD, mostly in sCJD and vCJD individuals (Orrú et al., 2014). The homozygosity at PRNP codon 129 was M/M and there were no mutations linked to gCJD. Summarizing the examination results with the typical appearances as progressive dementia, dystonia, myoclonus, psychiatric symptoms, limb tremor, cerebellar ataxia, and akinetic mutism, the patient was diagnosed with sCJD based on the criteria (Watson, et al., 2021). This diagnosis was subsequently confirmed by CDC. In the absence of specific therapeutic interventions, conservative care was administered in the local hospital. However, the patient experienced rapid disease progression and died within 20 days after returning to the local hospital (Puoti, et al., 2012).

## 3. Discussion

CJD is a transmissible neurodegenerative disease that usually leads to death within one year with symptoms such as cognitive and motor deficits (Puoti, et al., 2012). About 10–15% of cases are gCJD, which is associated with genetic mutation. sCJD occupies the majority of prion diseases which links to progressive neurodegeneration in the brain. The misfolded $PrP^C$ can cause further misfolding of normal $PrP^C$ in a cascade that results in neuronal death and associated clinical symptoms. The rapidly progressive cognitive decline is the predominant symptom of sCJD. The onset age of sCJD generally occurs at 70 years old (Uttley et al., 2020). Compared to other types, the pathogen of sCJD is unclear. The main pathological concept is an internal origin through a mutation in PRNP or the spontaneous misfolding of $PrP^C$ into $PrP^{Sc}$ (Orrú et al., 2014). sCJD is classified based on codon 129 polymorphism (M and V) and $PrP^{Sc}$ glycotype (1 and 2) into six subtypes (MM1, MM2, MV1, MV2, VV1 and VV2) (Puoti, et al., 2012). The subtypes MM1, MV1, and VV2 are the most common (Watson, et al., 2021; Puoti, et al., 2012).

The etiology of sCJD is not clear. The following cases who also diagnosed CJD suffered prior trauma (Harnish, et al., 2015; Nemani, et al., 2018; Scontrini, et al., 2015; Huang, et al., 2024). Some patients got CJD after implanting lyodura in the TBI surgery which had been reported in some cases (Iwasaki, et al., 2018; Preusser, et al., 2006). The relationship between CJD and traumatic event itself is not definite. The misdiagnosis between hydrocephalus and CJD has been reported as well (Uslu, et al, 2020). The thorough examination is important for the diagnosis and following treatment.

**Table 1**

Table summarizing the patients suffered from brain injury before.

| Author | Age | Gender | Injury etiology | Symptoms onset | MRI imaging | EEG | Survival time |
|---|---|---|---|---|---|---|---|
| Carissa etal, 2015 | 59y | Male | Gunshot | Confusion, delusional activity, difficulty focusing, and memory loss | Normal | Normal | 6m |
| Satish etal, 2018 | 84y | Male | Mild TBI | Cognitive and motor problems | Cerebral atrophy and small vessel white matter ischemic changes | NA | 5y |
| Satish etal, 2018 | 68y | Male | Mild TBI, Traffic event | Cognitive decline | Bilateral symmetric cortical restriction diffusion and FLAIR signal abnormality | NA | 4y |
| Satish etal, 2018 | 48y | Male | Traffic event | Bipolar disorder and PTSD | Asymmetric signal hyperintensity in parietal and occipital cortices, and caudate nucleus | Sharp wave periodic complexes | 7w |
| Alessandra etal, 2015 | 55y | Female | Fall | Walking unsteadiness | Traumatic lesions with bilateral hyperintensity of caudate and lenticular nuclei and slight cortical hyperintensity in the fronto-mesial regions | Diffused slowing | 5m (first symptom)/ 3m (post injury) |
| Brendan etal, 2024 | 54y | Male | Fall | Cognitive change | Gliosis and encephalomalacia in the right fronto-temporal region, restricted diffusion throughout the cortex, most pronounced in the left parietal lobe | NA | 1y |

Accurate diagnostic methods are important for subsequent treatment. Biopsy is mainly performed for accurate diagnosis and potential medical intervention (Watson, et al., 2021). MRI can be an effective non-invasive test for differential diagnosis (Watson, et al., 2021). DWI is the most sensitive method for diagnosing sCJD, typically showing asymmetric hyperintensity in at least three cortical non-contiguous gyri ("ribboning") or the striatum (caudate and rostral putamen) (Puoti, et al., 2012). DTI can quantitatively assess microstructural changes in normal brain tissue (Caverzasi, et al., 2014). DWI and FLAIR were recommended for diagnosis in 2009 (Watson, et al., 2021). CSF t-tau is elevated in many sCJD patients. The CSF 14-3-3 test, usually performed by western blot, has been a clinical diagnostic criterion for sCJD (Schmitz, et al., 2022). Compared with healthy individuals, expression of three miRNAs (hsa-let-7i-5p, hsa-miR-16-5p, and hsa-miR-93-5p) is markedly decreased (Norsworthy, et al., 2020). Plasma glial fibrillary acidic protein (pl-GFAP), associated with neuronal death, is elevated in sCJD patients (Bentivenga, et al., 2024). PrPSc is the characteristic lesion of sCJD. Clinical application of PrPSc amplification assays, such as protein misfolding cyclic amplification (PMCA) and real-time quaking-induced conversion (RT-QuIC), has greatly improved diagnostic accuracy. PMCA requires mixing patient brain samples or CSF with normal PrPC, followed by repeated steps to reveal PrPSc in suspected samples (Bentivenga, et al., 2024). RT-QuIC testing of CSF has high sensitivity (80-90%) (Jurcau, et al., 2024), and nasal brushing from the olfactory epithelium shows 97% sensitivity and 100% specificity (Wilson et al., 2019). Periodic sharp wave complexes (PSWCs) at approximately 1 cycle/second are typical CJD-like EEG findings (Puoti, et al., 2012). This patient exhibited characteristic symptoms consistent with sCJD.

The incidence of vCJD and iCJD has decreased due to prevention and control measures to remove prions from food and medical procedures. Understanding sCJD pathology may enable optimal treatment. Protein degradation therapies show potential to eliminate misfolded proteins and delay progression, though more studies are needed (Jurcau, et al., 2024).

The patient underwent multiple examinations for diagnosis. Thorough analysis enabled accurate diagnosis. Initially, we considered hydrocephalus due to ventricular enlargement. Brain atrophy can be difficult to differentiate from hydrocephalus as both show enlarged ventricles and similar symptoms. These conditions may coexist in one patient. Sometimes AD presents with ventricular enlargement and hydrocephalus, complicating diagnosis. sCJD incidence is lower than AD, making differential diagnosis between rare and common diseases challenging. The final diagnosis confirmed that thorough examination is essential.

## 4. Conclusion

The symptoms combined with clinical tests are important for the diagnosis of sCJD. The differential diagnosis of this disease from common diseases is difficult, especially when imaging is lacking. Though the diagnostic methods have made great progress, the effective treatment of sCJD still needs more research to confirm.

**Declaration of competing interest**

The authors declare that they have no known competing financial interests or personal relationships that could have appeared to influence the work reported in this paper.